# Behavioral response to strong aversive stimuli: A neurodynamical model

Kaushik Majumdar, Institute of Mathematical Sciences, CIT Campus, Taramani, Chennai – 600113, India.
E-mail: kaushik@imsc.res.in

**Abstract**

In this paper a theoretical model of functioning of a neural circuit during a behavioral response has been proposed. A neural circuit can be thought of as a directed multigraph whose each vertex is a neuron and each edge is a synapse. It has been assumed in this paper that the behavior of such circuits is manifested through the collective behavior of neurons belonging to that circuit. Behavioral information of each neuron is contained in the coefficients of the fast Fourier transform (FFT) over the output spike train. Those coefficients form a vector in a multidimensional vector space. Behavioral dynamics of a neuronal network in response to strong aversive stimuli has been studied in a vector space in which a suitable pseudometric has been defined. The neurodynamical model of network behavior has been formulated in terms of existing memory, synaptic plasticity and feelings. The model has an analogy in classical electrostatics, by which the notion of force and potential energy has been introduced. Since the model takes input from each neuron in a network and produces a behavior as the output, it would be extremely difficult or may even be impossible to implement. But with the help of the model a possible explanation for an hitherto unexplained neurological observation in human brain has been offered. The model is compatible with a recent model of sequential behavioral dynamics. The model is based on electrophysiology, but its relevance to hemodynamics has been outlined.

## 1. Introduction

Quick behavioral response to strong aversive stimuli (such as threat from a predator or an imminent danger of being hurt) is a key to survival throughout the animal kingdom. Network models of animal behavior have been elaborately discussed in (Schmajuk 1997). A neural network model of rats' anxiety behavior had been studied by Salum and colleagues (Salum et al. 2000), but they didn't take into account the functioning of the nerve cells in the brain during the task performance. Recently a neurodynamical model for conditional visuomotor association task has been proposed (Loh & Deco 2005). In this model a trial and error paradigm has been assumed in a stochastic decision space. An integrate and fire neuronal network model has been proposed to realize the paradigm. A neural network model of brain or cognitive state machine (CSM) to study decision making in a competitive environment has also been proposed (Rabinovich et al. 2006). The dynamics of making a choice from among multiple conflicting options has been formulated by Lotka-Volterra type of equations. It is evident that intelligent decisions in a sequential behavior have to be stable against noise and reproducible to allow memorization and reuse of successful decision sequences in the future. On the other hand, it also has to be sensitive to new information from the environment. These two fundamentally contradictory requirements have been taken care of in (Rabinovich et al. 2006).

In this paper a neurodynamical model of response behavior of a neural network to strong aversive stimuli has been presented. The assumption is that the network will behave in a manner to avoid repeating the negative experiences of the past. No specific neuronal network model has been assumed. In this approach information has been extracted (at least theoretically) from a behaving neuronal network by FFT on spike trains of all the neurons in the network, which should work across all networks of neurons irrespective of their architecture. The neurodynamical system has a unique representation in the information space where the Fourier coefficients of a spike train are arranged as a vector and uniquely represent the spike train. All the calculations have been carried out in that space. The model depends on past memory, synaptic plasticity and intensity of feeling. Interestingly, a short latency (120 – 160 ms) had been reported before response to aversive stimuli in the right prefrontal cortex of a human subject. No such latency was observed in case of pleasant or neutral stimuli (Kawasaki et al. 2001). The present model can offer a possible explanation for this apparently perplexing phenomenon. The following assumptions have been made:

1) Structure: Any behavior involves a network in the nervous system which if represented as a directed multigraph will have neurons as vertices and synapses as (directed) edges.
2) Function: The functional behavior of this network is manifested by the collective behavior of the neurons present in the network.
3) Memory: Memory of experiences of past behaviors of this network is stored entirely within the network and nowhere out side of it.
4) Plasticity: Plasticity of each synapse is a time dependent function (called *synaptic weight*) and the total plasticity of the network between any two behaviors is also a time dependent function called *network plasticity*.
5) Feeling: Feeling is associated with each behavior as a specific mathematical function which controls how experience associated with this behavior will mediate the intensity of effect of this behavior on any other behavior.
6) Interaction: Network plasticity mediates the interaction of the ongoing behavior with the memory of past behaviors stored in the network.

In section 2 each of the above points will be explained briefly and will be represented with appropriate mathematical expressions and equations. In section 3 with the help of an analogy from classical electrostatics the behavior dynamics of the network in terms of those expressions and equations will be formulated. Notion of a force like and a potential energy like expressions have been introduced. Despite computational difficulties involved with the model, in section 4 it has been related to reality first by offering an explanation to an hitherto unexplained observation in the human brain, second relating it with a successful decision making model and third relating this electrophysiological model to hemodynamical activity of the brain. The paper will be concluded with discussions and future directions.

**2. Modeling of the parts**

*A) Structure*

The closest computational analog of a neuronal circuit is a directed multigraph, whose each node will represent a neuron and each edge a synapse (Majumdar & Kozma, 2006). It can be represented as a three dimensional array $a[i,j,k]$. If there are $p$ different synapses joining the neuron $i$ with the neuron $j$ (assuming that all neurons in the brain have been numbered) then $a[i,j,k]$ will give the weight (the gain in synaptic transmission) of the $kth$ synapse joining the neuron $i$ with the neuron $j$, where $1 \leq k \leq p$. If the neuron $i$ and neuron $j$ are excitatory then $a[i,j,k]$ will be positive, if they are inhibitory then $a[i,j,k]$ will be negative. For given values of $i, j$ and $k$ $[i,j,k]$ uniquely determines a synapse and $a[i,j,k]$ represents the synaptic weight. In general $a[i,j,k]$ will be a function of time and may be written as $a[i,j,k](t)$ or $a_{ijk}(t)$, where $t$ denotes time.

*B) Function*

Computations in the central nervous system (CNS) have been viewed from three different angles – (a) synaptic computation (Abbott & Regehr 2004), (b) dendritic computation (London & Hausser 2005) and (c) neuronal computation (Koch 1999; Borisyuk & Rinzel 2005). In a neuronal network synapses are activating the neurons and neurons are activating the synapses. In this sense the aggregate behavior of the neurons in a network is equivalent to aggregate behavior of the synapses in that network. On an average every neuron receives input from about $10^4$ synapses and therefore from the computational modeling purpose it would be more convenient to consider the aggregate behavior of neurons rather than the aggregate behavior of synapses in a network. The behavior of a neuron during a particular epoch of time is completely represented by the spike train it generates during that period of time. A neuronal spike train carries information in the following manner (Kandel, et al. 2000).
  (1) The number of action potentials (spikes); and
  (2) The time intervals between them.
(Although this argument simplifies the description of brain functions to a great extent it ignores the reality of occurrence of change in the neural circuit without changing the firing patterns of the neurons. This has been discussed in Conclusion). The duration of a spike is typically 1 to 2 milliseconds (ms, depending on the temperature) (Koch 1999). When the sample frequency is high (1000 Hz or more) by FFT reliable information out of a neuronal spike train can be extracted. The FFT produces the vector $a_0 a_1 b_1 a_2 b_2 .... a_r b_r$, where

$$a_n = \frac{1}{2}(nth \text{ Fourier coefficient} + \text{conjugate of } nth \text{ Fourier coefficient}), \qquad (2.1)$$

$$b_n = \frac{1}{2i}(nth \text{ Fourier coefficient} - \text{conjugate of } nth \text{ Fourier coefficient}) \qquad (2.2)$$

$i = \sqrt{-1}$. For convenience it can be rewritten as $e(1).....e(2r+1)$, where

$$\left.\begin{array}{l} e(1) = a_0 \\ e(n) = a_{n/2} \quad \text{if } n \text{ is even} \\ e(n) = b_{2n-1} \quad \text{if } n \text{ is odd} \end{array}\right\}, \tag{2.3}$$

in uniform symbol. More conventionally the vector $v_k$ associated with the spike train of the *kth* neuron in the network can be written as

$$v_k = (e_k(1),...,e_k(2r+1))^T. \tag{2.4}$$

The suffix $T$ stands for transpose. If the duration of the spike train is $p$ seconds then $r = 500p$ (assuming sample frequency is 1000 Hz). Clearly the vector $v_k$ uniquely represents the spike train of the *kth* neuron in the network, assuming all the neurons in the network have been uniquely numbered. Since the Fourier series is convergent $e_k(2r+1) \to 0$ as $r \to \infty$.

Let there are $N$ neurons in the network. The behavior $B_i$ of the network for a duration of $p$ seconds is represented by the cluster of $N$ vectors $\{v_k^i\}_{k=1}^N$, where

$$v_k^i = (e_k^i(1),...,e_k^i(2r+1))^T, \tag{2.5}$$

$$r = \frac{f}{2}p, \tag{2.6}$$

$f$ is sample frequency. It is important to note that even when a neuron is oscillating below threshold for spike initiation, it can still release neurotransmitter and shape the final circuit output (Harris-Warrick & Marder 1991). But for simplicity of modeling I shall ignore this fact in this paper.

*C) Memory*

Memory is no single concept and there is no universally agreed upon definition of it. However the starting point for virtually any scientific analysis of memory involves a decomposition into processes of encoding, storage and retrieval (Schacter 2004). Wilder Penfield explored the cortical surface in more than a thousand epileptic patients. On rare occasions (about 8% of all the subjects he tried) he found that electrical stimulation in the temporal lobes produced a coherent recollection of an earlier experience (Kandel et al. 2000). A similar phenomenon has been observed by stimulating the inferotemporal (IT) cortex of macaque monkeys (Afraz et al. 2006). By mild stimulation to the IT in the macaque brain (previously trained to distinguish between face and non-face) impression of seeing a face was created in the mind of the animals where there was actually no face.

An opposite phenomenon has been reported for the human brain (Quian Quiroga et al. 2005). Visualization of images of interesting objects including faces of celebrities can make particular neurons to fire in the sub-region of the medial temporal lobes (MTL) of the human brain consisting of hippocampus, amygdala, entorhinal cortex and parahippocampal gyrus. This supports the hypothesis – object cognition activates a network in the brain and artificially activating the network to the appropriate degree will create the impression of perceiving the object in the brain even when it is not present in the environment. The following hypothesis seems to hold.

A particular cognition involves a particular network in the brain where the memory of the cognition remains stored. The higher processing areas of the network takes increasingly greater part in the cognition and also greater part in storing and retrieving the memory.

For the purpose of this paper it would be enough to be able to express memory in terms of behavior of a network. If the behavior of the network is $B_i$ the memory associated with it will be $M_i = \{u_k^i\}_{k=1}^N$. The relationship between $u_k^i$ and $v_k^i$ is determined by long term potentiation (LTP) and long term depression (LTD) if $M_i$ is residing in the network long after $B_i$ has happened. It may be appropriate to emphasize at this point that the collective firing pattern of neurons (i.e., the collective spike trains of neurons belonging to the network) to evoke $M_i$ is preserved by the synaptic connections in the network, for they control input to the neurons in response to the stimuli and therefore $M_i$ is stored in the synaptic strengths of the network. A metric between $B_i$ and $M_i$ can be defined in the following manner

$$d(B_i, M_i) = \frac{1}{N} \sqrt{\sum_{j=1}^{2r+1} \left[ \sum_{k=1}^{N} \left\{ e_k^{B_i}(j) - e_k^{M_i}(j) \right\} \right]^2} , \qquad (2.7)$$

where $N$ is the number of neurons in the network. Although virtually impossible to compute in this form $d$ can be a measure of plasticity of the whole network with respect to $B_i$ and $M_i$. Note that the time duration has been taken care of in $p$ at the time of determining $e_k^{B_i}(j)$ and $e_k^{M_i}(j)$ (neuronal firing patterns may not be identical at the time of a behavior and recalling it later both with respect to identical set of stimuli). A single network can mediate multiple behaviors (Harris-Warrick & Marder 1991) and therefore can store multiple memories.

Memory associated with a behavior may either be *positive* (appetitive) or *negative* (aversive) or *neutral* or a combination of them. The sense positive, negative and neutral are totally subjective and no attempt will be made here to define them. The meaning will become clear in the contexts in which they occur. A *simple behavior* is one with which only one type of memory (i.e., either positive or negative or neutral) remains associated. A behavior with combined types of memory can be called *complex behavior* and let us assume any behavior can be decomposed into simple behaviors. A network therefore can be thought to have memories of simple behaviors only. When the difference between $B_i$ and $M_i$ is small i.e., in (2.7) $d(B_i, M_i)$ becomes small, consolidation of $M_i$ is good. $d(B_i, M_i)$ is intimately related to synaptic plasticity.

*D) Plasticity*

(2.7) gives us an immediate measure (no matter however difficult it is or may even be impossible to implement) of plasticity of a neuronal network. The measure of combined long term plasticity $W(t)$ as a combination of long term potentiation (LTP) and long term depression (LTD) can be given by the following formula

$$W(p,t) = \frac{1}{m}\sum_{i=1}^{m} d(B_i, M_i), \tag{2.8}$$

where p is the duration of recording of the neural response when $B_i$ is happening and when $M_i$ is being retrieved both in response to identical set of stimuli and $t$ is the time difference between end of happening of $B_i$ and start of retrieving $M_i$. Usually $p$ remains fixed and therefore $W(p,t) = W(t)$. $m$ is the total number of past behaviors whose memory is still preserved in the network. When the gap between occurrence of $B_i$ and recalling of $M_i$ is long (30 minutes or more according to some estimate (Koch 1999)) (2.8) gives the long term plasticity and when it is shorter (2.8) gives short term plasticity. The following notion will also be useful

$$W_i(t) = d(B_i, M_i), \tag{2.9}$$

where $W_i(i)$ is the network plasticity between behavior $B_i$ and memory $M_i$ after time $t$.

*E) Feeling*

From a modeling or computational point of view the *feelings* may be taken as mediating the intensity of a behavioral response. In this sense if $B_i$ is a simple behavior the feeling associated with it determines how intensely negative or intensely positive will be the memory of it. Ideally $M_i$ should have an 'intensity distribution function' similar to a normal distribution function, which is decayed as the distance from the mean position is increased. $M_i$ is not a single point, but a cluster of points. So if the intensity distribution function is to be modeled after the normal distribution function the most appropriate candidate for the mean point turns out to be the mean point of $M_i$

$$\overline{M_i} = \frac{1}{N}\left(\sum_{k=1}^{N} e_k^{M_i}(1), \ldots, \sum_{k=1}^{N} e_k^{M_i}(2r+1)\right)^T, \tag{2.10}$$

where $T$ stands for transpose. The feeling function $F_i : R^{2r+1} \to R$ associated with $M_i$ is to be defined as

$$F_i(X) = \frac{1}{(2\pi)^{(2r+1)/2} |\Sigma|^{1/2}} \exp\left(-\frac{1}{2}(X - \overline{M_i})\Sigma^{-1}(X - \overline{M_i})\right), \tag{2.11}$$

where $\Sigma$ is the covariance matrix and $|\Sigma|$ is the determinant of $\Sigma$. Since $X$ has $2r+1$ independent component variables only diagonal entries of $\Sigma$ are nonzero and each of them is the variance of a component variable in $X$. $\Sigma$ is the parameter which controls the intensity of $F_i(X)$ beyond $\overline{M_i}$. The hormones and neuromodulators responsible for mediating feelings act by controlling the entries of $\Sigma$.

*F) Interaction*

Interaction means here the part played by synapses in the network in mediating the effects of $M_i$'s on an ongoing behavior $B$. In mathematical term it can be put as

$$I_i'(t) = f_i(W(p - t + t)) - f_i(W(p - t)), \tag{2.12}$$

where $I_i'(t)$ denotes the interaction and $t$ is any instant during occurrence of $B$. $f_i$ is a continuous function which is almost everywhere differentiable. When $B$ takes place the plasticity of the network changes and so also the interaction. Let us normalize interaction by the following formula

$$I_i(t) = \frac{\int_0^t I_0'(x)dx}{\int_0^T I_i'(x)dx}, \tag{2.13}$$

where $T$ is the duration of happening of $B$.

## 3. Integrative dynamics

Now let us briefly consider a phenomenon in electrostatics. Let there be seven charged particles on a plane and they are all in arbitrary but fixed positions. Four of them are positively charged and three are negative. A new negative charge is introduced, which is allowed to move freely in the plane, that is, it has two degrees of freedom along the X and Y axes. Assume apart from sign all the charges are quantitatively equal.

Let the locus of the introduced negative charge be $(x, y)$. The position of four positive charges be $\{(x_i, y_i)\}_{i=1}^{4}$ and that of the three negative charges be $\{(x_i, y_i)\}_{i=5}^{7}$. The Coulomb force $F(x, y)$ acting on the free charge is given by

$$f(x,y) = -\sum_{i=1}^{4} \frac{C}{(x-x_i)^2 + (y-y_i)^2} + \sum_{i=5}^{7} \frac{C}{(x-x_i)^2 + (y-y_i)^2}, \qquad (3.1)$$

where $C$ is a constant. (3.1) will govern the dynamics of the whole system. Now what should be the condition to keep the introduced charge fixed within a bounded region enclosing the seven fixed positioned charge so that the potential energy on the introduced particle becomes minimum?

On line of this analogy the dynamics of a new behavior $B$ should follow the governing expression $G(B)$ given by the following equation

$$G(B) = -\sum_{i=1}^{s} \frac{F_i(X) I_i(t)}{(d(M_i, B))^2} + \sum_{i=m-s+1}^{m} \frac{F_i(X) I_i(t)}{(d(M_i, B))^2}, \qquad (3.2)$$

where $m$ is the number of simple behaviors whose memories are stored in the network under consideration. It has been assumed that $B$ will be attracted towards $s$ behaviors with positive experience and will be repulsed by the remaining $m-s$ behaviors with negative experience. Like (3.1) there is no apparent reason why inverse square law should hold for (3.2) also. If the inverse square law does not hold then the denominators on the right side of (3.2) should be replaced by a general polynomial in the metric $d$, for any continuous function can be approximated to any desired degree within a compact interval by a suitable polynomial. Like the electrostatic situation for a given $M_i$ $G(M_i)$ must have a unique pole at the $\overline{M_i}$. This means in general (3.2) can have the form

$$G(B) = -\sum_{i=1}^{s} \frac{F_i(X) I_i(t)}{(d(M_i, B))^n} + \sum_{i=m-s+1}^{m} \frac{F_i(X) I_i(t)}{(d(M_i, B))^n}, \qquad (3.3)$$

for some positive $n$. However in this paper I shall adhere to (3.2), for the model here is essentially electrophysiological and therefore Coulomb interaction seems to be probable. In summary the dynamics of the new behavior $B$ is described by (2.7), (2.11), (2.13) and (3.2).

In analogy with the electrostatic system (3.2) is a 'force like' expression with which a 'potential energy like' expression needs to be associated, which for the sake of stability of the system (i.e., if perturbed infinitesimally will come back to the original state) needs to be minimized. Let the potential function associated with $B$ due to $M_i$ be $\phi^i$ given by

$$\phi^i = \sum_{k=1}^{N} \phi_k^i(e_k^i(1), \ldots, e_k^i(2r+1)). \qquad (3.4)$$

The potential function associated with the emerging new behavior $B$ due to all previous behaviors $B_i$ ($M_i$ is memory of $B_i$) be $\phi$, which is given by

$$\phi = \sum_{i=1}^{m} \phi^i. \tag{3.5}$$

To derive the potential energy classically from the force field the force must have to be conservative, i.e., independent of time (Goldstein 1950). In (3.2) the expression $I_i(t)$ is a function in time. However in case of a very strong aversive stimulus like, locating a predator dangerously close all of a sudden or being on the verge of falling down deep underneath from a very high roof top after a sudden slip, an extremely fast behavioral response must have to be shown. Within this duration synaptic plasticity does not get much chance to act and the feeling must have to be very strong (such as intense fear) to compensate for that as is evident from (3.2). This gives us an opportunity to treat $I(t)$ as a fixed quantity which makes $G(B)$ in (3.2) time independent for the duration of $B$. Then in analogy with the electrostatic system $G(B)$ must satisfy the following relation

$$G(B) = \frac{1}{N} \sum_{i=1}^{m} \sqrt{\sum_{k=1}^{N} \sum_{j=1}^{2r+1} \left\{ \frac{\partial \phi_k^i}{\partial (e_k^i(j))} \right\}^2}. \tag{3.6}$$

The value of $G(B)$ has only been considered and not its direction. (3.2) and (3.6) together give

$$\frac{1}{N} \sum_{i=1}^{m} \sqrt{\sum_{k=1}^{N} \sum_{j=1}^{2r+1} \left\{ \frac{\partial \phi_k^i}{\partial (e_k^i(j))} \right\}^2} = -\sum_{i=1}^{s} \frac{F_i(X) I_i(t)}{(d(M_i, B))^2} + \sum_{i=m-s+1}^{m} \frac{F_i(X) I_i(t)}{(d(M_i, B))^2}. \tag{3.7}$$

(3.7) describes the dynamics of the behavior of $B$ in the $R^{2r+1}$ under the assumption that $I_i(t)$ is constant (otherwise the 'force field' would not have been conserved and only space dependent potential expressions could not have been brought in the dynamics). Also the time scale is very small. $\phi$ as given by (3.5) will have to be minimized (minimization of energy is an important criterion for neural computation (Laughlin 2004)), which means

$$\frac{\partial \phi}{\partial (e_k^i(j))} = 0, \forall j \in \{1, \ldots, (2r+1)\}. \tag{3.8}$$

Combining (3.4) and (3.5) the expression for $\phi$ becomes

$$\phi = \sum_{i=1}^{m}\sum_{k=1}^{N} \phi_k^i(e_k^i(1),......,e_k^i(2r+1)). \tag{3.9}$$

If (3.9) is to give a global minimum it should not only be true when $\phi$ is given by (3.9), but also it must hold when $\phi$ is given by

$$\phi = \sum_{i=1}^{m} x_i \sum_{k=1}^{N} y_k \phi_k^i(e_k^i(1),......,e_k^i(2r+1)), \tag{3.10}$$

where $x_i, y_k \in (1-c, 1+c)$ for some small $c > 0$. This is because a global minimum is a very stable position and therefore under a small perturbation the system always comes back to it. Let $x_i y_k = z_{ik}$. Then combining (3.8) and (3.10) the following is obtained

$$\begin{pmatrix} a_{1,1,1} & a_{1,2,1} & \cdot & \cdot & a_{m,N,1} \\ a_{1,1,2} & a_{1,2,2} & \cdot & \cdot & a_{m,N,2} \\ \cdot & \cdot & \cdot & \cdot & \cdot \\ \cdot & \cdot & \cdot & \cdot & \cdot \\ \cdot & \cdot & \cdot & \cdot & \cdot \\ \cdot & \cdot & \cdot & \cdot & \cdot \\ \cdot & \cdot & \cdot & \cdot & \cdot \\ \cdot & \cdot & \cdot & \cdot & \cdot \\ a_{1,1,2r+1} & a_{1,2,2r+1} & \cdot & \cdot & a_{m,N,2r+1} \end{pmatrix} \begin{pmatrix} z_{1,1} \\ z_{1,2} \\ \cdot \\ \cdot \\ \cdot \\ \cdot \\ \cdot \\ \cdot \\ z_{m,N} \end{pmatrix} = \begin{pmatrix} 0 \\ 0 \\ \cdot \\ \cdot \\ \cdot \\ \cdot \\ \cdot \\ \cdot \\ 0 \end{pmatrix}, \tag{3.11}$$

where $a_{i,k,j} = \dfrac{\partial \phi_k^i(e_k^i(1),......,e_k^i(2r+1))}{\partial(e_k^i(j))}$. Note that each $z_{i,k}$ or $z_{ik}$ can take uncountably infinitely many values from some (small) open interval and for all of them (3.11) holds under the perturbation principle. Therefore the $(2r+1) \times mN$ matrix on the left of (3.11) must represent a null linear transformation and it must be a null matrix, which implies

$$a_{i,k,j} = \frac{\partial \phi_k^i(e_k^i(1),......,e_k^i(2r+1))}{\partial(e_k^i(j))} = 0, \forall i,k,j. \tag{3.12}$$

(3.7) and (3.12) together imply

$$\sum_{i=1}^{s} \frac{F_i(X)I_i(t)}{(d(M_i,B))^2} = \sum_{i=m-s+1}^{m} \frac{F_i(X)I_i(t)}{(d(M_i,B))^2}. \tag{3.13}$$

(3.13) says, "Negative and positive experiences in a neuronal network must counter balance each other for a stable (which will not be altered due to presence of some amount of noise or distraction) unsupervised learning from the experience gained through a new behavior."

**4. Application**

What do we get from the principle enunciated at the end of the last section in response to a strong aversive stimulus? Its mathematical formulation (3.13) says, in the face of strong aversive stimuli (such as an imminent danger) the ensuing behavior $B$ must avoid repeating the past behaviors $B_i$'s with negative experience $M_i$'s. Note that the stimuli can invoke an $M_i$ if and only if $B_i$ is at least partially activated by the stimuli. (3.13) says the ensuing behavior $B$ will have to be such that none of the $B_i$'s with negative experience is repeated. This means in the information space $R^{2r+1}$ $B$ will have to sit away from each $M_i^-$, which denotes memory of a negative behavior. $B$ will have its own positive and negative parts which will later be stored as new positive and negative memories in the network. $B$ cannot be neutral. This will be the subject of a future work.

*A) Aversion response latency*

The prefrontal cortex participates in linking perception of stimuli to the guidance of behavior including the flexible execution of strategies for obtaining rewards and avoiding punishments as an organism interacts with its environment. Recording from neurons within healthy tissue in the ventral sites of the right prefrontal cortex short latency (120–160ms) responses selective for aversive visual stimuli have been observed. No such latency was observed for pleasant or neutral stimuli (Kawasaki et al. 2001). (3.13) can offer us an explanation of this phenomenon. Aversive stimuli do evoke negative (or aversive) memory (that is how the stimuli are identified as aversive, even a novel aversive stimulus will have to be decomposable into known aversive features) and therefore the ensuing behavior $B$ must ensure that it has minimum overlap with the negative memories in the space $R^{2r+1}$. This in turn makes sure that $B$ acts on the network as less as possible to repeat the behaviors associated with the negative memories.

When there will be only aversive stimuli (no pleasant stimulus) the new behavior will have to be organized to avoid the stimuli particularly when the stimuli are strongly aversive (like an immediate threat). Clearly a strongly aversive stimulus must invoke the memory $M_j^-$ of at least one simple negative behavior with which a strong feeling is to be associated. In other words amygdale make sure that the $\Sigma$ in (2.11) have larger entries. Then there must be the memory $M_j^+$ of a simple positive behavior which can take appropriate action in response to the behavior (sensation) corresponding to $M_j^-$. Even a new born is hardwired to express displeasure by crying in response to an aversive stimulus so that some one else (possibly the mother) is alerted and come in help to avoid the stimulus. In order to avoid the aversive stimuli the dynamics of $B$ will be such that (3.13) can take the following form

$$\sum_{j=1}^{k}\frac{F_j(X)I_j(t)}{\left(d(M_j^+,B)\right)^2} = \sum_{i=1}^{m}\frac{F_i(X)I_i(t)}{\left(d(M_i,B)\right)^2}, \qquad (4.1)$$

where $F_j(X)$ is the feeling associated with $M_j^+$ and $I_j(t)$ is the network interaction between $B$ and $M_j^+$, $k$ is the number of simple positive behavior recalled. This will counterbalance the aversive effect of the stimuli and make sure that (3.13) holds.

Whereas in case of pleasant or neutral stimuli (without the presence of aversive stimuli) avoidance is not necessary, the brain is free to repeat the behaviors with pleasant memory. In that case no global minimization of the potential function will be necessary and left side of (3.7) does not need to be zero. Therefore equality in (4.1) will also not be necessary. The new behavior can sit anywhere and no 'optimum positioning' (making the potential function globally minimum) will be necessary and this will require less time for a response and therefore there will be no latency in the response to pleasant stimuli.

When there will be both positive and negative stimuli (3.13) will hold. Thus it appears that whenever aversive (negative) stimuli are present either (3.13) or (4.1) needs to be calculated. Time taken by this 'calculation' in the brain is the reason for latency when there are aversive stimuli. Whereas no such latency is necessary for positive or neutral stimuli. Probably the brain does not calculate the way shown in this paper. Had it done so it would have taken a much longer latency. But the reasoning here is compelling enough to conclude that calculations responsible for the observed latency do take place in the brain in one form or the other.

*B) Sequential decision making*

A brain always has to make choices i.e., a behavior is a series of switching or decision making procedures. Rabinovich and colleagues have considered the dynamics of decision making (DM) by the brain or cognitive state machine (CSM) at the psychological level (Rabinovich et al. 2006). Whereas the focus of this paper is on the dynamics of neural substrate of such processes. At each DM instant of a CSM the underlying neural dynamics of the DM generates the psychological process of DM in the CSM.

The dynamics of the CSM is governed by the Lotka-Volterra type equation (Rabinovich et al. 2006)

$$\dot{a}_i = a_i\left[\sigma_i(I,t) - \left(a_i + \sum_{j\neq i}^{N}\rho_{ij}a_j\right)\right] + \eta_i(t), \qquad (4.2)$$

where $a_i(t)$ is the state of the CSM, $\sigma_i(I,t)$ is a control function which controls the dynamics given by (4.3), $I$ is the input (environmental stimulus), $\rho_{ij}$ is a coupling constant between $a_i$ and $a_j$ based on genetic and memorized information (very similar to the interaction $I_i(t)$ (2.13) between the memories of the old behavior and the ensuing new behavior, which in this paper has been taken to be constant), $N$ the total number of states and $\eta_i(t)$ is the external noise. Note that in (4.2) the meaning of $\sigma_i$ and $N$ are different than anywhere else in the earlier part of this paper.

$$\tau\dot{\sigma}_i = -\frac{\partial U_i(\sigma_i, I)}{\partial \sigma_i}, \qquad (4.3)$$

where $U_i$ is a potential function and $\tau$ is characteristic time which is very small. Difference in states occur at the $t_i$ where a decision needs to be made as shown in Figure 1. $a_i$ can be chosen at an instant $t_j$, where a decision has to be made, as many ways as there are minima of $U_i$.

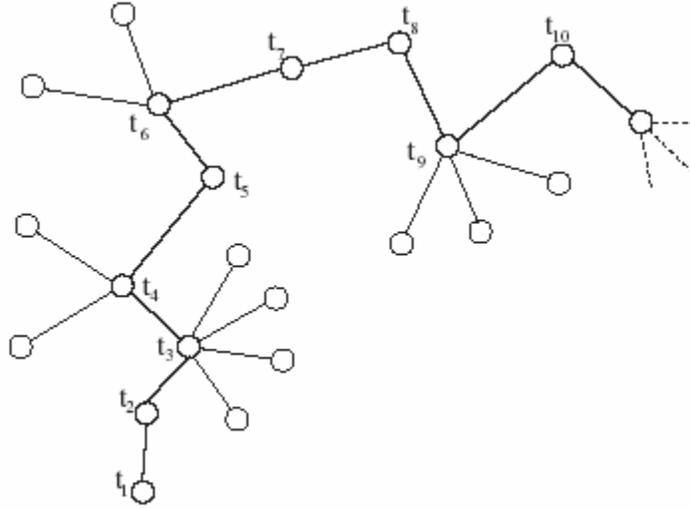

Figure 1: A sequence of cognitive states. Thin lines are possible paths and the decision path has been shown by thick lines. $t_3, t_4, t_6, t_9$ are the instances to make a decision. Adopted from (Rabinovich et al. 2006).

(3.13) is concerned about the dynamics of neural substrate of each single decision making (DM) in the sequence of DM's shown in Figure 1. (3.13) is also derived by minimizing the potential function (3.9). Since the dynamics given by (3.7) is concerned about a local decision making in Figure 1, it was possible to take global optimum of the potential function given by (3.9) (global optimum of local potential function) and its stability ensures that the local dynamics is not perturbed by a small amount of noise. It is very important for stable cognition or stable behavior. According to this model the alteration in behavior (to make it dynamic) should be introduced by changing feeling (given by (2.11)) and synaptic plasticity (given by (2.13)), which happens from state to state in a CSM. Therefore despite the global minimum of the local dynamics the CSM manages to move on from one state to the next till it reaches the end of life which may be signified by lack of emotional impetus. On the other hand at the sequential DM level in a CSM (Figure 1) introduction of noise plays a significant role in dynamically altering the behavior (equation (4.2)). In both the dynamical systems of this paper and that of Rabinovich et al. the time length has been taken to be short. The duration has been taken so short that the dynamics of $\sigma_i(I,t)$ (in (4.3)) and $I_i(t)$ (in (2.13)) do not change within that period. In this scenario comparing (3.13) and (4.2) it appears that the external noise $\eta_i(t)$ may take an important role in mediating feeling (emotional distraction or attraction with respect to an external stimulus).

(3.13) and (4.1) describe how new behavior is formed depending on synaptic plasticity, memory and feeling. However there is no deterministic way to compute the new behavior from (3.13) or (4.1) even if all the information are available. Also (3.13) or (4.1) will be extremely difficult (if not impossible) to compute except for cases like a simple behavior (such as inking or gill withdrawal) of a simple animal (such as the sea snail *Aplysia californica*). This is in conformity with the fact that closer is the model to the neuronal network level of the brain the more difficult it would be to implement. On the other hand the CSM dynamics given by (4.2) is very convenient to compute.

*C) Potential function*

The potential function introduced by (3.9) in the neuronal network dynamics of the brain can be related to metabolic energy consumption during activation of the network to execute a behavioral or cognitive task. (3.9) is based on electrophysiology and metabolism is related to hemodynamics. Lot of research is going on to find the relationship between these two activities of the brain (Logothetis & Wandell 2004). Recently a correlation study between stimulus based ERP and fMRI in different parts of the human brain has been reported (Gore et al. 2006). Despite convincing evidence of their interdependence, such as in the form of dependence of fMRI on ERP or EEG, precise knowledge about the nature of dependence is still lacking. A linear transform model had been proposed based on the hypothesis that fMRI responses are proportional to local average neural activity averaged over a period of time (Boynton et al. 1996). It has been reported that in the visual cortex of macaque monkey fMRI response depends closely on the local field potential (LFP) (Logothetis et al. 2001).

Note that the potential function given by (3.5) can be obtained from (3.7) when the time course is short and the quantities on the right side are known. From (3.7) it is clear that the potential function $\phi$ (which is defined on the Fourier coefficients of neural spike trains (3.9)) depends on past memory, synaptic plasticity and feelings. In case of a potential function $\phi$ sum of local components will give the function over the whole space, and therefore no matter how large and distributed the network is, local component of $\phi$ at a point $x$ on the cortex (let be denoted by $\phi_x$) can be represented by (3.7) where only the neurons in a small neighborhood of $x$, denoted by $n(x)$, will participate in the computation of $\phi_x$. Since $\phi_x$ defined on the Fourier coefficients of spike trains of the neurons in $n(x)$ and the value of $\phi_x$ depends on past memories, plasticity of synapses within $n(x)$ and part of feeling arising within $n(x)$ it is supposed to have a role in BOLD fMRI response at $x$ with respect to a stimulus. $\phi_x$ may have an affine or linear relation with stimulus driven difference in local BOLD signal at $x$ as modeled in (Logothetis & Wandell 2004). But this needs experimentations to establish.

**Conclusion**

In this paper a system level model of the brain functions for behavior has been proposed. The model needs input from individual neurons under the assumption that a brain circuit responsible for a behavior can be understood by the behavior of neurons alone. This assumption has limitations, for significant change in a neural network may occur at the spiking sub-threshold level in the neurons and therefore without changing the spike trains (Harris-Warrick & Marder 1991). Note that FFT based information retrieveal technique followed in this paper will hold equally good for subthreshold signals, whereas spike detection or prediction algorithms may not work for those signals. To account for a fuller neural computation synaptic computation (Abbott & Regehr 2004) and dendritic computation (London & Hausser 2005) will also have to be incorporated. A close investigation into the nature of $I(t)$ (equation (2.13)) will inevitably call attention to synaptic and dendritic computations. The model equation (3.2) will be valid in all generality, but the important consequence (3.13) cannot be drawn as easily as shown in this paper.

The effectiveness of the model dynamics is only for a very short period of time and therefore only behaviors in response to strong aversive stimuli have been considered, for such response behaviors must have to be very quick. The time duration is so short that it has been assumed – the average change in the collection of synapses in a network remains fixed within that time. This is an important constraint for introducing potential energy function whose global minimization gives stability to the behavioral response in the sense that it remains unperturbed in a noisy environment.

Even without incorporating synaptic and dendritic computations the model stands extremely difficult to implement (for example, calculating $B_i$ and then calculating $M_i$ will be very challenging). However it can be implemented in case of a simple behavior by a simple animal. In invertebrates like the sea snail *Aplysia* lesser number of neurons will have to be monitored (only a few in case of simple behaviors like gill withdrawal or inking) and single cell recordings will be less challenging than the mammals. This will make the verification of the system possible. In case of human brain functions if the behavior of the circuit involved can be monitored to a good extent by recording signals only from a few neurons verification of the model may be feasible for some very simple behavior like the following. Consider the experiment of tapping the leg (Kandel et al. 2000). If there is a sharp edge such that dragging the leg too much backward will make it bump on the sharp edge which is another aversive stimulus. The final position of the leg will be away from the source of tapping as well as the sharp edge.

If a relationship between the potential function $\phi$ and the metabolic or hemodynamic activity of the brain can be established, study of the system will become easier. Introduction of an electrophysiological potential function in the neural computation is a significant outcome of the FFT based information extraction method followed in this paper. Apart from the prospect of relating it to metabolic energy requirements of the brain $\phi$ gives the opportunity of deriving interesting theoretical results as shown in this paper.

Since the expression of feeling as given by (2.11) is independent of time here the detail of feeling did not have to be considered. But feeling is likely to be dependent on time in the long run and in that case the entries of $\Sigma$ (which is a diagonal matrix) will be function of time and the dynamics will be much more complicated. How the eigen values of $\Sigma$ are controlled by the limbic system of the brain is an open question. No attempt has been made in this paper to answer the question.

**Acknowledgement**

The author thankfully acknowledges the Institute of Mathematical Sciences in Chennai, India for a postdoctoral fellowship under which this work has been carried out. Some comments by Peter Dayan on a preliminary version of this manuscript are also being acknowledged.